\documentclass[twocolumn,preprintnumbers,showpacs,pre,floatfix,amsmath,amssymb,notitlepage,superscriptaddress]{revtex4-1}
\usepackage{epsfig}
\usepackage{subfigure}
\usepackage{amsmath}
\usepackage{color}
\usepackage{amssymb}
\usepackage{setspace}
\usepackage{graphicx}
\usepackage{dcolumn}
\usepackage{bm}
\usepackage{times}
\usepackage{enumerate}
\usepackage{footnote}

\input epsf

\graphicspath{{figures/}}

\begin{document}

\author{Per Sebastian Skardal}
\email{persebastian.skardal@trincoll.edu} 
\affiliation{Department of Mathematics, Trinity College, Hartford, CT 06106, USA}

\author{Alex Arenas}
\email{alexandre.arenas@urv.cat} 
\affiliation{Department d'Enginyeria Inform\'{a}tica i Matem\'{a}tiques, Universitat Rovira i Virgili, 43007 Tarragona, Spain}

\title{Higher-order interactions in complex networks of phase oscillators promote abrupt synchronization switching}


\begin{abstract}
Synchronization processes play critical roles in the functionality of a wide range of both natural and man-made systems. Recent work in physics and neuroscience highlights the importance of higher-order interactions between dynamical units, i.e., three- and four-way interactions in addition to pairwise interactions, and their role in shaping collective behavior. Here we show that higher-order interactions between coupled phase oscillators, encoded microscopically in a simplicial complex, give rise to added nonlinearity in the macroscopic system dynamics that induces abrupt synchronization transitions via hysteresis and bistability of synchronized and incoherent states. Moreover, these higher-order interactions can stabilize strongly synchronized states even when the pairwise coupling is repulsive. These findings reveal a self-organized phenomenon that may be responsible for the rapid switching to synchronization in many biological and other systems that exhibit synchronization without the need of particular correlation mechanisms between the oscillators and the topological structure.
\end{abstract}


\maketitle

The collective dynamics of network-coupled dynamical systems has been a major subject of research in the physics community during the last decades~\cite{Strogatz2003,Arenas2008PhysRep,Dorfler13,Pikovsky2015} due to a wide range of applications applications including cardiac rhythms~\cite{Karma2013Rev}, power grid dynamics~\cite{Rohden2012PRL}, and proper cell circuit behavior~\cite{Prindle2011Nature}. In particular, our understanding of both natural and man-made systems has significantly improved by studying how network structures and dynamical processes combine to shape overall system behaviors. This interplay gives rise to novel nonlinear phenomena like switch-like abrupt transitions to synchronization~\cite{GomezGardenes2011PRL,Skardal2019PRL,Rev2019} and cluster states~\cite{Pecora2014NatComms,Cho2017PRL}. Recent work in physics and neuroscience have specifically highlighted the importance of higher-order interactions between dynamical units, i.e., three- and four-way interactions in addition to pairwise interactions, and their role in shaping collective behavior~\cite{Petri2014Interface,Giusti2016JCN,Ashwin2016PhysD,Reimann2017,Otter2017EPJDS,millan18,Sizemore2018JCN,Leon2019PRE}, prompting the network science community to turn its attention to higher-order structures to better represent the kinds of interactions that one can find beyond typical pairwise interactions~\cite{Horak2009}. These higher-order interactions are often encoded in simplicial complexes~\cite{Salnikov2019EJP} that describe the different kinds of simplex structure present in the network: a filled clique of $m+1$ nodes is known as an $m$-simplex, and together a set of 1-simplexes (links), 2-simplexes (filled triangles), etc. comprise the simplicial complex. While simplicial complexes have been proven to be very useful for analysis and computation in high dimensional data sets, e.g., using persistent homologies~\cite{Otter2017EPJDS}, little is understood about their role in shaping dynamical processes, save for a handful of examples~\cite{Schaub2019,Iacopini2019NatComms,Matamalas2019}. 

In parallel to the previous developments, there has been also a lot of attention on another phenomena related to the collective dynamics of network-coupled oscillators namely the explosive synchronization phenomenon, see~\cite{Rev2019} and references therein. Explosive synchronization consists of an abrupt switch between incoherent and synchronized states, that can be achieved by the interplay between the network structure and the oscillators dynamics, being the most simple prescription that of each oscillator having a natural frequency proportional to the number of connections in the network. This mathematical finding is becoming particularly important in neuroscience, where bistability and fast switching of states are very relevant to understand, bistable perception~\cite{Wang2013}, epileptic seizures in the brain~\cite{Andrew1989,Wang2017}, or hypersensitivity in chronic pain of Fibromyalgia patients~\cite{Lee2018}. However, the mechanisms for this abrupt switching to happen are still unclear. The specificities of the networks should not be the most relevant parameter, given that human wiring is not equivalent between individuals~\cite{Finn2015}, and then we rely on another aspect, the higher-order (beyond pairwise) interactions in the network. The collective dynamics of sources and loads in large-scale power grids provides another important application where abrupt synchronization transitions play an important role~\cite{Motter2013NatPhys}.

Motivated by the above mentioned dynamical processes, here we study the dynamics of heterogeneous phase oscillators with higher-order interactions on simplicial complexes with 1-, 2-, and 3-simplex interactions. Similarly to our recent approach in~\cite{Skardal2019PRL}, where we investigate the desynchronization phenomena taking in to account 2-simplexes, here we aim to understand the effect of higher-order interactions that combine 1-, 2-, and 3-simplex interactions in the emergence of synchronization. In this previous study we showed that although 2-simplex interactions alone did not lead to any synchronization transition (i.e., they do not destabilize the incoherent state) the synchronization, they do give rise to abrupt desynchronization transitions. Here we show that the combination of multiple higher-order interactions gives rise to both abrupt synchronization and desynchronization transitions, allowing the system to easily switch between incoherent and synchronized states with relatively small changes to system parameters.
We use the celebrated Kuramoto model~\cite{Kuramoto1984} to scrutinize the higher order dynamics in complex networks. Previous studies already revealed a rich phase diagram where nonpairwise interactions are considered, showing multi-stability~\cite{Tanaka2011}, quasiperiodicity~\cite{Rosenblum2007}, and even chaos~\cite{Bick2016}. Our contribution aligns with these previous works and demonstrates that higher-order interactions provide a natural mechanism for the emergence of explosive synchronization.

For a simplicial complex of $N$ nodes we propose an extension of the Kuramoto--Sakaguchi phase rotator model~\cite{Sakaguchi1986} on networks to the higher-order Kuramoto model whose equations of motion are given by
\begin{align}
\dot{\theta}_i=\omega_i&+\frac{K_1}{\langle k^1\rangle}\sum_{j=1}^NA_{ij}\sin(\theta_j-\theta_i)\nonumber\\
&+\frac{K_2}{2\langle k^2\rangle}\sum_{j=1}^N\sum_{l=1}^NB_{ijl}\sin(2\theta_j-\theta_l-\theta_i)\nonumber\\
&+ \frac{K_3}{6\langle k^3\rangle}\sum_{j=1}^N\sum_{l=1}^N\sum_{m=1}^NC_{ijlm}\sin(\theta_j+\theta_l-\theta_m-\theta_i),\label{eq:01}
\end{align}
where $\theta_i$ is the phase of oscillator $i$, $\omega_i$ is its natural frequency (typically assumed to be drawn from a distribution $g(\omega)$), and $K_1$, $K_2$, and $K_3$ are the coupling strengths of 1-, 2-, and 3-simplex interactions, respectively. Importantly, these addition forms of coupling (i.e., those with $K_2$ and $K_3$ coefficients) come directly from higher-order terms that emerge from phase-reductions of limit-cycle oscillators~\cite{Ashwin2016PhysD,Leon2019PRE}. The network structure (assumed to be undirected and unweighted) is encoded in the 1-simplex adjacency matrix $A$, 2-simplex adjacency tensor $B$, and 3-simplex adjacency tensor $C$, where $A_{ij}=1$ if nodes $i$ and $j$ are connected by a link (and otherwise $A_{ij}=0$), $B_{ijl}=1$ if nodes $i$, $j$, and $l$ belong to a common 2-simplex (and otherwise $B_{ijl}=0$), and $C_{ijlm}=1$ if nodes $i$, $j$, $l$, and $m$ belong to a common 3-simplex (and otherwise $C_{ijlm}=0$). For each node $i$ we denote the $q$-simplex degree $k_i^q$ as the number of distinct $q$-simplexes node $i$ is a part of, and $\langle k^q\rangle$ is the mean $q$-simplex degree across the network. (Note that each division by $\langle k^q\rangle$ in equation~(\ref{eq:01}) amounts to a rescaling of the respective coupling strength).

\begin{figure}[t]
\centering
\includegraphics[width=0.8\linewidth ]{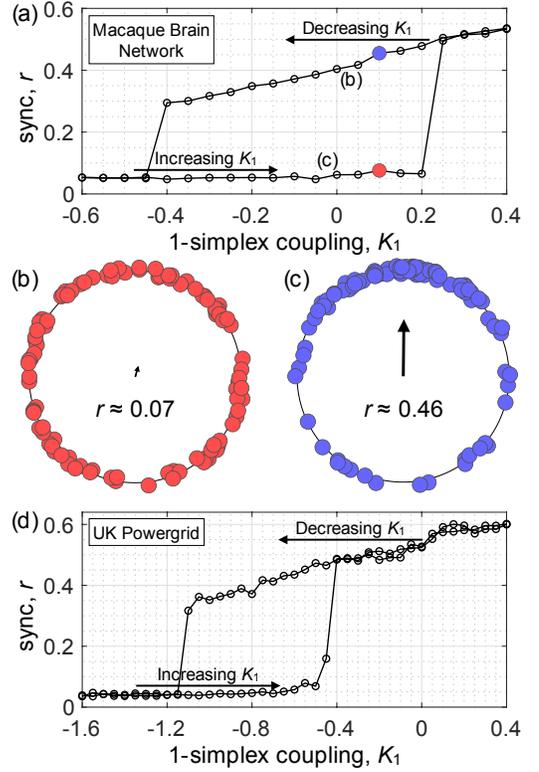}
\caption{{\bf Abrupt synchronization in simplicial complexes: Macaque brain and UK power grid networks.} (a) The synchronization profile describing the macroscopic system state by the order parameter $r$ as a function of 1-simplex coupling $K_1$ for higher-order coupling strengths $K_{2}=1.6$ and $K_3=1.1$ using the Macaque brain network. Results are obtained by adiabatically increasing $K_1$ from $-0.6$ to $0.4$, then subsequently decreasing $K_1$ from $0.4$ back to $-0.6$. This protocol reveals a hysteresis loop with abrupt synchronization and desynchronization transitions at $K_1^{\text{sync}}\approx0.25$ and $K_{1}^{\text{desync}}\approx-0.4$ with a bistable region of incoherence and synchronization in between. Incoherent and synchronized states at $K_1=0.1$ are illustrated in panels (b) and (c), respectively. (d) The synchronization profile as in (a) using the UK power grid network and higher-order coupling strengths $K_{2}=2.2$ and $K_3=3.3$.} \label{fig1}
\end{figure}

\begin{figure*}[t]
\centering
\includegraphics[width=0.7\linewidth ]{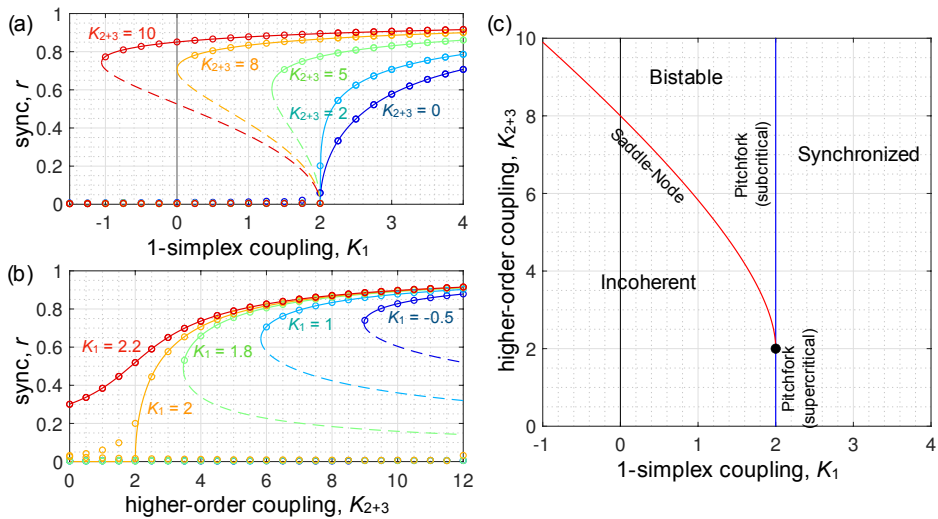}
\caption{{\bf Abrupt synchronization in simplicial complexes: All-to-all case.} Synchronization profiles describing the macroscopic system state: (a) the order parameter $r$ as a function of 1-simplex coupling $K_1$ for higher-order coupling $K_{2+3}=0$, $2$, $5$, $8$, and $10$ (blue to red) and (b) the order parameter $r$ as a function of higher-order coupling $K_{2+3}$ for 1-simplex coupling $K_1=-0.5$, $1$, $1.8$, $2$, and $2.2$. Solid and dashed curves represent stable and unstable solutions given by equation~(\ref{eq:05}), respectively, and circles denote results taken from direct simulations of equation~(\ref{eq:02}) with $N=10^4$ oscillators. (c) The full stability diagram describing incoherent, synchronized, and bistable states as a function of 1-simplex coupling $K_1$ and higher-order coupling $K_{2+3}$. Blue and red curves correspond to pitchfork and saddle-node bifurcations, which collide at a codimension-two point (black circle) at $(K_1,K_{2+3})=(2,2)$. For $K_{2+3}<2$ and $K_{2+3}>2$ the pitchfork bifurcation is supercritical and subcritical, respectively.} \label{fig2}
\end{figure*}

Taking inspiration from the importance of simplicial complexes in the brain, which displays rich synchronization dynamics~\cite{Schnitzler2005Nature}, we consider as a motivating example the dynamics of equation~(\ref{eq:01}) on the Macaque brain dataset which consists of $242$ interconnected regions of the brain~\cite{Amunts2013Science}. The adjacency matrix $A$ is taken to be undirected and 2- and 3-simplex structures are constructed by identifying each distinct triangle and tetrahedron from the 1-simplex structures. The 2- and 3-simplex coupling strengths are fixed to $K_2=1.6$ and $K_3=1.1$ as the 1-simplex coupling strength is varied and natural frequencies are drawn identically and independently from the standard normal distribution. In Fig.~\ref{fig1}(a) we plot the amplitude $r$ of the complex order parameter $z=re^{i\psi}=N^{-1}\sum_{j=1}^Ne^{i\theta_j}$ as $K_1$ is first increased adiabatically from $K_1=-0.6$ to $0.4$, then decreased. These simulations reveal that the presence of higher-order interactions in simplicial complexes give rise to abrupt (a.k.a. explosive) synchronization transitions~\cite{GomezGardenes2011PRL}, as the system quickly transitions from the incoherent state ($r\approx0$) to a partially synchronized state ($r\sim1$) at $K_1^{\text{sync}}\approx 0.25$ as $K_1$ is increased, then another abrupt transition from synchronization to incoherence occurs at $K_1^{\text{desync}}\approx -0.4$ as $K_1$ is decreased. For $K_1\in[K_1^{\text{desync}},K_1^{\text{sync}}]$ the system admits a bistability where both incoherent and synchronized states are stable. In Figs.~\ref{fig1}(b) and (c) we highlight this bistabiliy by showing the incoherent and synchronized states, respectively, for $K_1=0.1$, illustrating for $40\%$ of the oscillators (chosen randomly) placed appropriately on the unit circle with their respective order parameter values $r\approx0.07$ and $0.46$.

The results presented above illustrate two new critical findings using a real brain dataset. First, the presence of higher-order interactions, i.e., 2- and 3-simplexes, can induce abrupt synchronization transitions without any additional dynamical or structural ingredients. Incoherent and synchronized states have been mapped to resting and active states of the brain~\cite{Deco2013Trends}, respectively, with abrupt transitions representing quick and efficient mechanisms for switching cognitive tasks. However, previous work has shown that in the presence of only 1-simplex coupling, properties such as time-delays~\cite{Lee2009PRL} or degree-frequency correlations~\cite{GomezGardenes2011PRL} are needed to induce such transitions. Second, the presence of higher-order interactions can create and stabilize a synchronized state even when 1-simplex coupling is negative, i.e., repulsive. Thus, higher-order interactions nonlinear effects that support synchronization on the macroscopic scale. To emphasize the broader implications of this finding, we plot in Fig.~\ref{fig1}(d) the synchronization profile of the order parameter $r$ vs 1-simplex coupling $K_1$ (again both increasing and decreasing $K_1$ to highlight the explosive transitions and bistability) for higher-order coupling strengths $K_{2}=2.2$ and $K_3=3.3$ on the UK power grid network~\cite{Rohden2012PRL}. Here, since the network is strongly geometric and adjacent nodes are geographically close to one another, and therefore likely similarly affected by local events, we identify 2-simplexes as 3-paths (i.e., paths of three connected nodes, a.k.a., wedges) and 3-simplexes as 4-paths and 4-stars (i.e., three nodes all connected to a fourth central node). The qualitatively similar behavior displayed here demonstrates a wide range of important synchronization applications where higher-order interactions may significantly affect the dynamics.

To better understand the dynamics that emerge in the system above, we turn our focus to a population of all-to-all coupled oscillators. The governing equations, which also serves as the mean-field approximation for equation~(\ref{eq:01}), is given by
\begin{align}
\dot{\theta}_i=\omega_i&+\frac{K_1}{N}\sum_{j=1}^N\sin(\theta_j-\theta_i)+\frac{K_2}{N^2}\sum_{j=1}^N\sum_{l=1}^N\sin(2\theta_j-\theta_l-\theta_i)\nonumber\\
&+ \frac{K_3}{N^3}\sum_{j=1}^N\sum_{l=1}^N\sum_{m=1}^N\sin(\theta_j+\theta_l-\theta_m-\theta_i).\label{eq:02}
\end{align}
In the all-to-all case given by equation~(\ref{eq:02}) the system can be treated using the dimensionality reduction of Ott and Antonsen~\cite{Ott2008Chaos}, yielding a low dimensional system that governs the macroscopic dynamics via the order parameter $z=re^{i\psi}$. In particular, by considering the continuum limit of infinitely-many oscillators and applying the Ott-Antonsen ansatz (see Methods for details), we obtain for the amplitude $r$ and angle $\psi$ the simple differential equations
\begin{align}
\dot{r} &= -r + \frac{K_1}{2}r(1-r^2)+\frac{K_{2+3}}{2}r^3(1-r^2),\label{eq:03}\\
\dot{\psi} &=\omega_0,\label{eq:04}
\end{align}
where we have assumed that the natural frequency distribution $g(\omega)$ is Lorentzian with mean $\omega_0$ and the new coupling strength is given by the sum of the 2- and 3-simplex coupling strengths, i.e., $K_{2+3}=K_2+K_3$. Note first that the amplitude and angle dynamics of $r$ and $\psi$ completely decouple and that the angle dynamics evolve with a constant angular velocity equal to the mean of the frequency distribution. Thus, by entering an appropriate rotating frame and shifting initial conditions we may set $\psi=0$ without any loss of generality. Moreover, the higher-order interactions, i.e., 2- and 3-simplexes mediated by the coupling strength $K_{2+3}$, surface in the form of cubic and quintic nonlinear terms. This implies that the stability of the incoherent state, given by $r=0$, (which is always an equilibrium) is not affected by the higher-order interactions. However, these nonlinear terms that originate from the higher-order interactions mediate the possibility of synchronized states. In particular, one or two synchronized states also exists, given by
\begin{align}
r=\sqrt{\frac{K_{2+3}-K_1\pm\sqrt{(K_1+K_{2+3})^2-8K_{2+3}}}{2K_{2+3}}},\label{eq:05}
\end{align}
where the plus and minus signs correspond to stable and unstable solutions when they exist.

We now show that the all-to-all case illustrates, in an analytically tractable setting, all the novel dynamics observed in the Macaque example (see Fig.~\ref{fig1}). First, in Fig.~\ref{fig2}(a) we plot steady-state solutions of the order parameter $r$ as a function of the 1-simplex coupling strength $K_1$ for a variety of higher-order coupling strengths $K_{2+3}=0$, $2$, $5$, $8$, and $10$ (blue to red). Analytical predictions given by equation~\ref{eq:05} are plotted as solid and dashed curves (for stable and unstable branches, respectively), and circles represent results from direct simulation of equation~(\ref{eq:01}) with $N=10^4$ oscillators. For sufficiently small higher-order coupling (e.g., $K_{2+3}=0$) the transition to synchronization is second-order, occurring via a supercritical pitchfork bifurcation. However, as $K_{2+3}$ is increased through a critical value of $K_1^{\text{sync}}=2$ the synchronized branch folds over itself, giving rise to hysteresis and abrupt transitions between incoherence and synchronization for larger values of higher-order coupling (e.g., $K_{2+3}=5$, $8$, and $10$). In this regime the pitchfork bifurcation at $K_1^{\text{sync}}=2$ becomes subcritical and a saddle-node bifurcation emerges at a lower value of $K_1$, denoted $K_1^{\text{desync}}$, where the synchronized branch first appears. These two bifurcations correspond to the abrupt transitions observed in Fig.~\ref{fig1}. We also observe that for $K_{2+3}\ge8$ the synchronized branch stretches into the negative region $K_1<0$ (e.g., $K_{2+3}=10$), again demonstrating that higher-order interactions can stabilize synchronized states even when pairwise interactions are repulsive. In Fig.~\ref{fig2}(b) we plot similar results as the higher-order coupling strength $K_{2+3}$ is varied for a variety of 1-simplex coupling strengths, $K_{1}=-0.5$, $1$, $1.8$, $2$, and $2.2$ (blue to red). These curves highlight the existence and absence of bistability for $K_{1}<2$ and $K_1>2$, respectively. In Fig.~\ref{fig2}(c) we provide the full stability diagram for the system, denoting the pitchfork bifurcations at $K_{1}^{\text{sync}}=2$ (supercritical and subcritical for $K_{2+3}<2$ and $K_{2+3}>3$) in blue and the saddle-node bifurcation, given by $K_{1}^{\text{desync}}=2\sqrt{2K_{2+3}}-K_{2+3}$, in red. The region bounded by these curves corresponds to bistability between synchronization and incoherence, and is born at the intersection between the two bifurcations at the codimension-two point $(K_1,K_{2+3})=(2,2)$.

\begin{figure}[t]
\centering
\includegraphics[width=0.8\linewidth ]{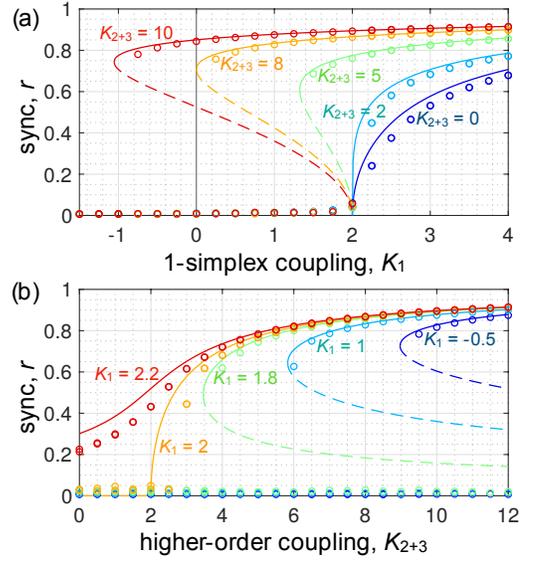}
\caption{{\bf Synchronization in the multiplex simplicial complex model.} For the multiplex model of simplicial complexes, synchronization profiles describing the macroscopic system state: (a) the order parameter $r$ as a function of 1-simplex coupling $K_1$ for higher-order coupling $K_{2+3}=0$, $2$, $5$, $8$, and $10$ (blue to red) and (b) the order parameter $r$ as a function of higher-order coupling $K_{2+3}$ for 1-simplex coupling $K_1=-0.5$, $1$, $1.8$, $2$, and $2.2$. Circles represent direct simulations on a network of $N=10^4$ nodes with mean degrees $\langle k^1\rangle=\langle k^2\rangle=\langle k^3\rangle=30$ and solid and dashed curves represent stable and unstable solutions of the mean field approximation given by equation~(\ref{eq:05}).} \label{fig3}
\end{figure}

Having demonstrated the novel synchronization dynamics that arise from higher-order interactions in simplicial complexes in a real brain dataset and the all-to-all scenario, we lastly turn to a synthetic network example, constructing a simplicial complex via a three-layer multiplex, where the $q^{\text{th}}$ layer consists of $q$-simplexes. In particular, aiming for such a multiplex with mean degrees $\langle k^1\rangle$, $\langle k^2\rangle$, and $\langle k^3\rangle$, we construct each layer randomly, placing $M_1=N\langle k^1\rangle/2$ 1-simplexes (i.e., links) in the first layer, $M_2=N\langle k^2\rangle/3$ 2-simplexes (i.e., filled triangles) in the second layer, and $M_3=N\langle k^3\rangle/4$ 3-simplexes (i.e., filled tetrahedra) in the third layer. (Note that the first layer is a classical Erd\H{o}s-R\'{e}nyi network~\cite{Erdos1960} and the second and third layers are the generic extensions using 2- and 3-simplexes instead of typical links.) In Figs.~\ref{fig3}(a) and (b) we plot the the order parameter $r$ vs 1-simplex coupling $K_1$ and higher-order coupling $K_{2+3}$, resecptively, for a multiplex network of $N=10^4$ oscillators with mean degrees $\langle k^1\rangle=\langle k^2\rangle=\langle k^3\rangle=30$ in circles. Similar to Figs.~\ref{fig2}(a) and (b), solid and dashed curves represent the analytical results for the mean-field approximation from the all-to-all case. These results illustrate that the mean-field approximation accurately describes the dynamics of such randomly generated simplicial complexes.

The results presented above demonstrate that higher-order interactions in networks of coupled oscillators, which are encoded on the microscopic scale of by a simplicial complex, give rise to added nonlinearities in the macroscopic system dynamics. These nonlinearities give rise to two new phenomena that are not present in the absence of higher-order interactions, i.e., when interactions are solely pairwise. First, these nonlinearities induce abrupt transitions between incoherent and synchronized states without additional characteristics such as time delays or network-dynamics correlations. In the context of brain dynamics, incoherent and synchronized states correspond to resting and active states, with abrupt transitions facilitating efficient switching between cognitive tasks~\cite{Deco2013Trends}. Second, when nonlinearities are sufficiently strong they create and stabilize synchronized states even when pair-wise coupling is repulsive. Thus, even as certain kinds of coupling may degrade over time due to synaptic plasticity, the presence of other kinds of coupling may be enough to sustain bistability regimes between incoherence and synchronization. We note that in this paper we have taken brain dynamics as our primary motivating example due to the existence of direct evidence of higher-order interactions in a system with synchronization properties~\cite{Petri2014Interface,Giusti2016JCN,Reimann2017,Sizemore2018JCN}. However, more general results suggest that higher-order interactions may be important in broader classes of physical systems~\cite{Ashwin2016PhysD,Leon2019PRE} including large-scale power grids~\cite{Skardal2015SciAdv}, indicating that the nonlinear phenomena observed in this context may point to other novel behaviors that arise from such interactions in different contexts.

{\it Methods.} Here we detail the dimensionality reduction used to derive equations~(\ref{eq:02}) and (\ref{eq:03}) We begin by rewriting equation~(\ref{eq:01}) using the complex order parameters $z$ and $z_2=N^{-1}\sum_{j=1}^Ne^{2i\theta_j}$, yielding
\begin{align}
\dot{\theta}_i=\omega_i+\frac{1}{2i}\left(He^{-i\theta_i}-H^*e^{i\theta_i}\right),\label{eq:M01}
\end{align}
where $H=K_1z+K_2z_2z^*+K_3z^2z^*$ and $*$ denotes the complex conjugate. In the thermodynamic limit we may represent the state of the system using the density function $f(\theta,\omega,t)$, where $f(\theta,\omega,t)d\theta d\omega$ gives the fraction of oscillator with phase in $[\theta,\theta+d\theta)$ and frequency in $[\omega,\omega+d\omega)$ at time $t$. Because oscillators are conserved and frequencies are fixed, $f$ satisfies the continuity equation
\begin{align}
0=\frac{\partial f}{\partial t}+\frac{\partial}{\partial \theta}\left\{f\left[\omega_i+\frac{1}{2i}\left(He^{-i\theta_i}-H^*e^{i\theta_i}\right)\right]\right\}.\label{eq:M02}
\end{align}
Expanding $f$ into its Fourier series $f(\theta,\omega,t)=\frac{g(\omega)}{2\pi}\left[1+\sum_{n=1}^\infty\hat{f}_n(\omega,t)e^{in\theta}+\text{c.c.}\right]$ (where c.c. denoted the complex conjugate of the previous term), we follow Ott and Antonsen~\cite{Ott2008Chaos} hypothesis that Fourier coefficients decay geometrically, i.e., $\hat{f}_n(\omega,t)=\alpha^n(\omega,t)$ for some function $\alpha$ that is analytic in the complex $\omega$ plane. Remarkably, after inserting this ansatz into $f$ and $f$ into equation~(\ref{eq:M02}), all Fourier modes collapse onto the same constraint for $\alpha$, giving the single differential equation
\begin{align}
\dot{\alpha}=-i\omega\alpha+\frac{1}{2}\left(H^*-H\alpha^2\right).\label{eq:M03}
\end{align}
Moreover, in the thermodynamic limit we have that $z^*=\iint f(\theta,\omega,t)e^{i\theta}d\theta d\omega=\int\alpha(\omega,t)g(\omega)d\omega$. By letting $g$ be Lorentzian with mean $\omega_0$ and width $\Delta$, i.e., $g(\omega)=\Delta/{\pi[\Delta^2+(\omega-\omega_0)^2]}$, this integral can be evaluated by closing the contour with the infinite-radius semi-circle in the negative-half complex plane and using Cauchy's integral theorem~\cite{Ablowitz2003}, yielding $z^*=\alpha(\omega_0-i\Delta,t)$. (Similarly, we have that $z_2^*=\alpha^2(\omega_0-i\Delta)=z^{*2}$.) Evaluating equation~(\ref{eq:M03}) at $\omega=\omega_0-i\Delta$ and taking a complex conjugate then yields
\begin{align}
\dot{z}&=-\Delta z+i\omega_0 z \nonumber\\
&+\frac{1}{2}\left[\left(K_1z+K_{2+3}z^2z^*\right)-\left(K_1z^*+K_{2+3}z^{*2}z\right)z^2\right].\label{eq:M04}
\end{align}
Using the rescaled time $\hat{t}=\delta t$ and rescaled coupling strengths $\hat{K}_1=K_1/\Delta$ and $\hat{K}_{2+3}=K_{2+3}/\Delta$ (effectively setting $\Delta=1$) and separating equation~(\ref{eq:M04}) into evolution equations for $r$ and $\psi$ yields (after dropping the $\wedge$-notation) equations~(\ref{eq:02}) and (\ref{eq:03}). Note that in the particular case in which $K_2=K_3$ equation~(\ref{eq:M04}) contains a second harmonic in the phase differences, encapsulating and in accordance with previous results in the literature~\cite{Filatrella07,Vlasov15}. 




\bibliographystyle{plain}

\end{document}